\documentclass{PoS}

 \usepackage{graphics}
 \usepackage{graphicx}
 \usepackage{epsfig}
 \usepackage{longtable}
 \usepackage{latexsym}
 \usepackage{amsmath}
 \usepackage{amsthm}
 \usepackage{amsfonts}
 \usepackage{amssymb}
 \usepackage{dsfont}
 \usepackage{bm}
 \usepackage{graphics,psfrag}
 \usepackage{graphicx,psfrag}
 \usepackage{subfigure}
 \graphicspath{{}}

 \newcommand{\be}{\begin{equation}}
 \newcommand{\ee}{\end{equation}}
 \newcommand{\bey}{\begin{eqnarray}}
 \newcommand{\eey}{\end{eqnarray}}

 \newcommand{\DCI}{\ensuremath{D_\srm{CI}} }

 \newcommand{\srm}[1]{\textrm{\scriptsize{#1}}}
 
 \newcommand{\E}{\mathrm{e}}
 \newcommand{\FC}{\;,}
 
 \renewcommand{\imath}{\mathrm{i}}

\title{Excited meson spectroscopy with two chirally improved quarks}
\ShortTitle{Excited meson spectroscopy with two chirally improved quarks}

\author{\speaker{Georg P.~Engel}
\\
        Institut f\"ur Physik, FB Theoretische Physik, Universit\"at
Graz, A--8010 Graz, Austria\\
        E-mail: \email{georg.engel@uni-graz.at}
}
\author{C.~B.~Lang\\
        Institut f\"ur Physik, FB Theoretische Physik, Universit\"at
Graz, A--8010 Graz, Austria\\
        E-mail: \email{christian.lang@uni-graz.at}
}
\author{Markus Limmer\\
        Institut f\"ur Physik, FB Theoretische Physik, Universit\"at
Graz, A--8010 Graz, Austria\\
        E-mail: \email{markus.limmer@uni-graz.at}
}
\author{Daniel Mohler\\
        TRIUMF, 4004 Wesbrook Mall Vancouver, BC V6T 2A3, Canada \\
        E-mail: \email{mohler@triumf.ca}
}
\author{Andreas Sch\"afer\\
        Institut f\"ur Theoretische Physik, Universit\"at
Regensburg, D--93040 Regensburg, Germany\\
        E-mail: \email{andreas.schaefer@physik.uni-regensburg.de}
}

\abstract{
The excited isovector meson spectrum is explored using two chirally 
improved dynamical quarks.
Seven ensembles, with pion masses down to $\approx$ 250 MeV are discussed 
and used for extrapolations to the physical point.
Strange mesons are investigated using partially quenched s-quarks.
Using the variational method, we extract excited states in several 
channels and most of the results are in good agreement with experiment.
}

\FullConference{XXIX International Symposium on Lattice Field Theory \\
		 July 10 - 16 2011\\
		 Squaw Valley, Lake Tahoe, California}

\begin{document}
%%%%%%%%%%%%%%%%%%%%%%%%%%%%%%%%%%%%%%%%%%%%%%%%%%%%%%%%%%%%%%%%%%%%%%%%%%%%%%%

%%%%%%%%%%%%%%%%%%%%%%%%%%%%%%%%%%%%%%%%%%%%%%%%%%%%%%%%%%%%%%%%%%%%%%%%%%%%%%%
\section{Introduction}
\label{sec:intro}
%%%%%%%%%%%%%%%%%%%%%%%%%%%%%%%%%%%%%%%%%%%%%%%%%%%%%%%%%%%%%%%%%%%%%%%%%%%%%%%
\noindent
The bulk of experimental knowledge about QCD observables is contained in the hadron spectrum, listed by the 
Particle Data Group \cite{Nakamura:2010zzi}. 
So far, lattice QCD is the only known technique to perform corresponding ab-initio calculations.
We use the Chirally Improved Dirac operator \cite{Gattringer:2000js}, an approximate solution
of the Ginsparg-Wilson (GW) equation \cite{Ginsparg:1981bj}.
We present results for ground and excited states of the isovector mesons in 2-flavor QCD, making use of the variational method. 
We access strange mesons by including partially quenched strange quarks in the analysis.
Recent preceeding work was published in \cite{Engel:2009cq,Engel:2010my,Engel:2010dy}.
A more extensive discussion of the results will be found in \cite{Engel:2011}.

%%%%%%%%%%%%%%%%%%%%%%%%%%%%%%%%%%%%%%%%%%%%%%%%%%%%%%%%%%%%%%%%%%%%%%%%%%%%%%%
\section{Simulation details}
\label{sec:simdet}
%%%%%%%%%%%%%%%%%%%%%%%%%%%%%%%%%%%%%%%%%%%%%%%%%%%%%%%%%%%%%%%%%%%%%%%%%%%%%%%
\noindent
The simulation method is detailed in \cite{Gattringer:2008vj}. 
The Chirally Improved Dirac operator ($\DCI$) is obtained by plugging the most general
ansatz for a Dirac operator into the GW equation and comparing the coefficients. 
Furthermore, one level of stout smearing enters the definition of $\DCI$, and the L\"uscher-Weisz gauge action is used. 
We generate the dynamical configurations with a Hybrid Monte-Carlo (HMC) algorithm.
We simulate seven ensembles with lattices of size $16^3 \times 32$, for details see Table \ref{tab:simdet}. 
\begin{table*}[b]
\begin{center}
\begin{tabular}{cccccccc}
\hline
\hline
set	& $m_\pi$[MeV]	& $a$[fm]&	$\beta_{LW}$&$m_0$&   $m_{\pi}L$& configs \\
\hline
A50&	596(7)          & 0.1324(11)&        4.70&  -0.050& 6.40&200 \\
A66&	255(7)          & 0.1324(11)&        4.70&  -0.066& 2.72&200 \\
B60&	516(6)          & 0.1366(15)&        4.65&  -0.060& 5.72&300 \\
B70&	305(6)          & 0.1366(15)&        4.65&  -0.070& 3.38&200 \\
C64&	588(6)          & 0.1398(14)&        4.58&  -0.064& 6.67&200 \\
C72&	451(5)          & 0.1398(14)&        4.58&  -0.072& 5.11&200 \\
C77&	330(5)          & 0.1398(14)&        4.58&  -0.077& 3.74&300 \\
\hline
\end{tabular}
\end{center}
\caption{Parameters used in the simulation of the seven ensembles.
Given are the pion mass $m_\pi$, the lattice spacing $a$, the gauge 
coupling parameters $\beta_{LW}$, the quark mass parameters $m_0$, the 
factor $m_{\pi}L$ with the spatial extent $L$ of the lattice 
and the number of (approximately independent) configurations.
The lattice spacing is set using the static potential assuming a Sommer 
parameter of $r_{0,exp} = 0.48\,$ fm, detailed in the text.
}
\label{tab:simdet}
\end{table*}
The variational method \cite{Michael:1985ne,Luscher:1990ck} is applied to extract ground and excited states.
The correlation matrix for a given set of interpolators is
\begin{eqnarray}
C_{ij}(t) 		&=& 	\langle 0 \vert O_i(t)  O_j^{\dag} \vert 0\rangle  = \sum_{k=1}^N \langle 0 \vert O_i \vert k\rangle \langle k \vert O_j^{\dag} \vert 0\rangle\,\E^{-t E_k}.
\label{eq:varmeth1}
\end{eqnarray}
The idea of the variational method is to offer a basis of suitable interpolators, from which the system
chooses the linear combinations closest to the physical eigenstates $|k\rangle$.
The generalized eigenvalue equation
\begin{eqnarray}
C(t)\, \vec{v}_k 	&=& 		\lambda_k(t,t_0)\, C(t_0)\, \vec{v}_k \FC \qquad \lambda_k(t,t_0) 	\propto 	\E^{-(t-t_0)\,E_k} \left( 1+\mathcal{O}(\E^{-(t-t_0)\,\Delta E_k}) \right)\ ,
\label{eq:varmeth2}
\end{eqnarray}
gives the energies of the eigenstates, where $\Delta E_k$ is the difference between 
$E_k$ and its closest energy level. 
The eigenvectors represent the linear combinations of the used set of 
interpolators which are close to the physical states related to the eigenvalues.
We use two Gaussian (narrow and wide) and derivative sources.
The Gaussian sources are obtained using gauge-covariant Jacobi smearing, 
the derivative source by applying the covariant derivative on the wide source. 
Combining these quark sources, several interpolators are constructed in each hadron channel.
All sources are located in single time slices, built on configurations  which are hypercubic-smeared (HYP) in the spatial directions three times.
Complete tables of interpolators will be found in the 
Appendix of \cite{Engel:2011} and differ slightly  from \cite{Engel:2010my}.
We consider isovector-mesons, which are free of disconnected diagrams. 
Within our present framework, isoscalars can be discussed only with the systematic error of neglected disconnected diagrams.
We use the meson interpolator construction as described in \cite{Gattringer:2008be}, 
which is similar to constructions previously used in \cite{Lacock:1996vy,Liao:2002rj,Dudek:2007wv}. 

In order to set the scale, the dimensionless lattice observable $a/r_0$ (cf. Table \ref{tab:simdet}) is extrapolated linearly in $(am_\pi)^2$ for ensembles sharing the same gauge coupling.
We extrapolate to the physical value of the dimensionless product $m_\pi r_0$ (represented by a parabola in this specific graph).
This procedure defines a scale for each gauge coupling, independent of the particular quark mass parameters, where physical input is used only at the physical point.

%%%%%%%%%%%%%%%%%%%%%%%%%%%%%%%%%%%%%%%%%%%%%%%%%%%%%%%%%%%%%%%%%%%%%%%%%%%%%%%
\section{Results}
\label{sec:results}
%%%%%%%%%%%%%%%%%%%%%%%%%%%%%%%%%%%%%%%%%%%%%%%%%%%%%%%%%%%%%%%%%%%%%%%%%%%%%%%
\noindent

%%%%%%%%%%%%%%%%%%%%%%%%%%%%%%%%%%%%%%%%%%%%%%%%%%%%%%%%%%%%%%%%%%%%%%%%%%%%%%%
\subsection{The light 0$^{-+}$ channel: $\pi$}
\begin{figure}
\begin{minipage}[t!]{74mm}
\includegraphics[width=73mm,clip]{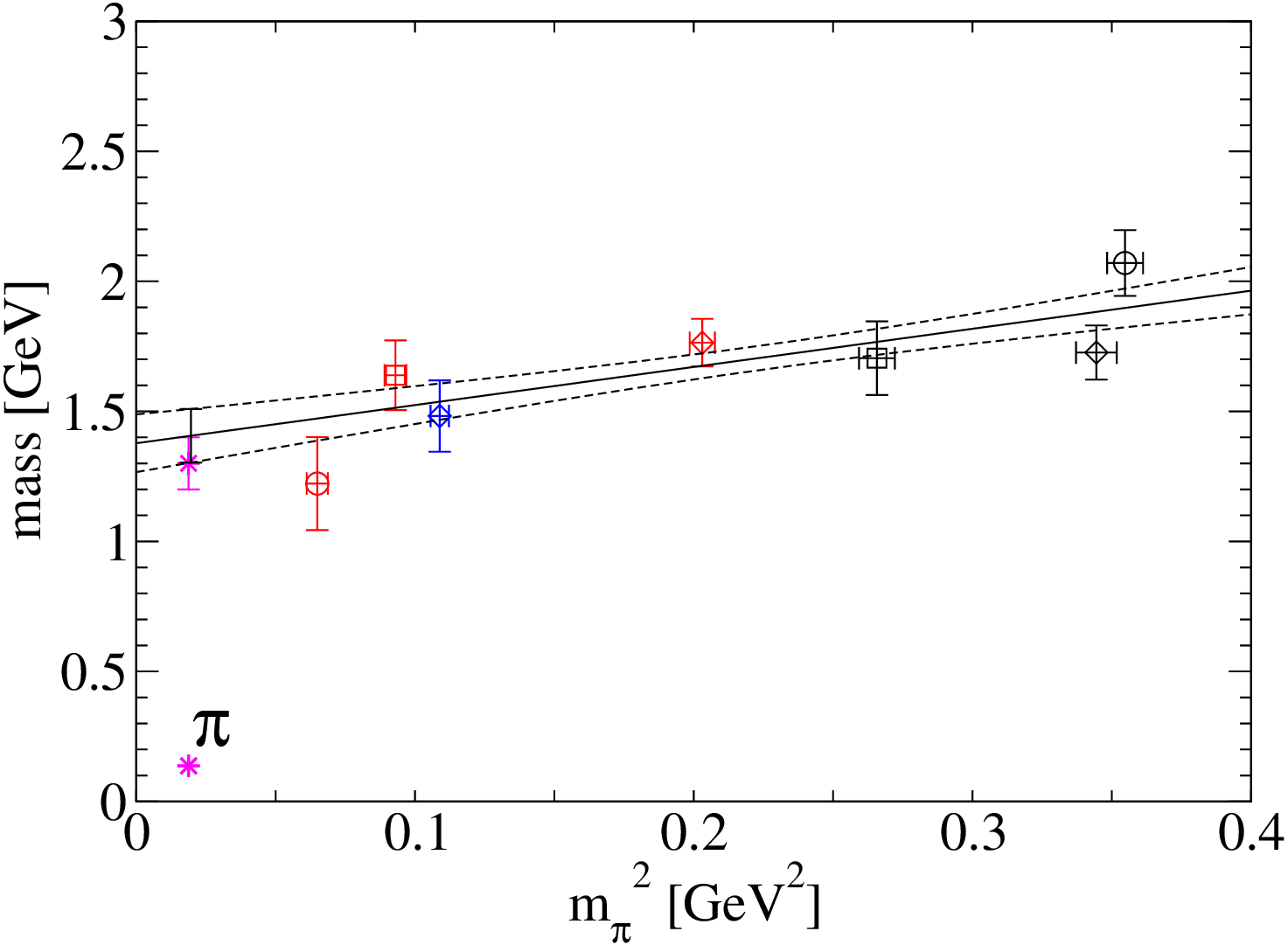}
\caption{
Mass plot for the $0^{-+}$ channel ($\pi$), first excitation.
The ground state defines the abscissa.
}
\label{fig:pi}
\end{minipage}
\hfill
\begin{minipage}[t!]{74mm}
\vspace*{-0.5mm}
\includegraphics[width=73mm,clip]{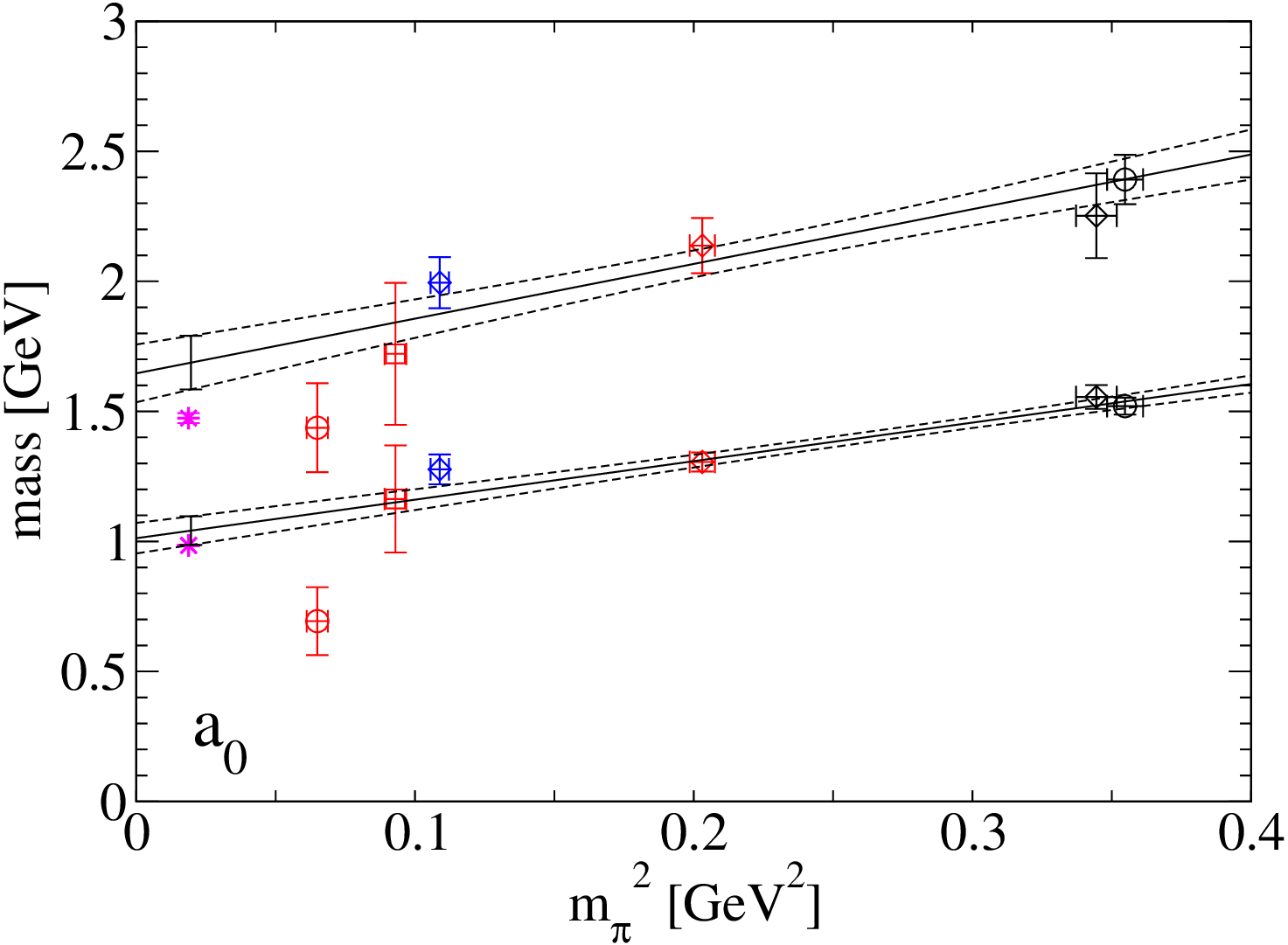}
\caption{
Mass plot for the 0$^{++}$ channel ($a_0$), ground state and first excitation.
}
\label{fig:a0}
\end{minipage}
\end{figure}
\noindent
As usual, the pion ground state mass is used as scale for the chiral extrapolation for all other observables.
For the first excitation we use the set of operators (1,2,17).
Due to the finiteness of the lattice, the backrunning pion limits the possible fit range for the first excitation, in particular at small pion masses (see Fig.~\ref{fig:pi}).
Nevertheless, masses can be extracted and the chiral extrapolation hits the experimental $\pi(1300)$ within $1\sigma$ (see Fig.~\ref{fig:pi}).

%%%%%%%%%%%%%%%%%%%%%%%%%%%%%%%%%%%%%%%%%%%%%%%%%%%%%%%%%%%%%%%%%%%%%%%%%%%%%%%
\subsection{The light 0$^{++}$ channel: $a_0$}
\noindent
In the 0$^{++}$ channel, we extract the ground state and the first excited state using the set of interpolators (10,12,13).
The energy levels of ensemble A66 appear unexpectedly light  (see Fig.~\ref{fig:a0}), 
which may be related to scattering states becoming visible in the analysis at light sea quark masses.
Thus, we cannot exclude sizeable contributions from the two-particle state $\pi \eta_{2}$.
However, the chiral extrapolations -- dominated by the larger pion mass data -- 
are compatible with the experimental 
one-particle states $a_{0}(980)$ within 1$\sigma$ and with $a_{0}(1450)$ within 1$\sigma$.
Due to the comparatively small value of $m_\pi L$ for A66 there may be a 
significant finite size effect, as well.

%%%%%%%%%%%%%%%%%%%%%%%%%%%%%%%%%%%%%%%%%%%%%%%%%%%%%%%%%%%%%%%%%%%%%%%%%%%%%%%
\subsection{The strange 0$^{-}$ channel: $K$}
\begin{figure}
\begin{minipage}[t!]{74mm}
\includegraphics[width=73mm,clip]{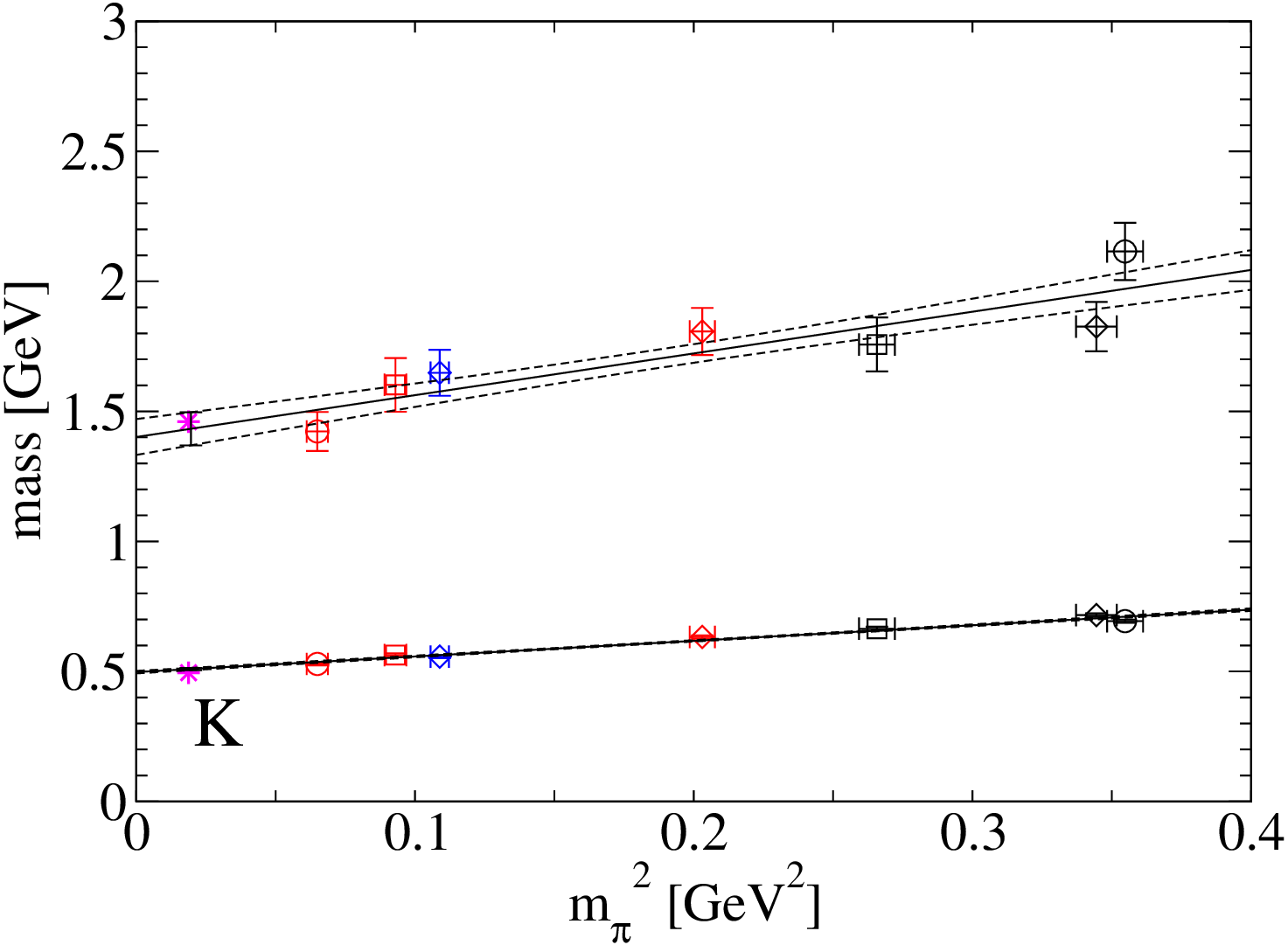}
\caption{
Mass plot for the strange 0$^{-}$ channel ($K$), ground state and first excitation.
}
\label{fig:K}
\end{minipage}\hfill
\begin{minipage}[t!]{74mm}
\vspace*{-0.5mm}
\includegraphics[width=73mm,clip]{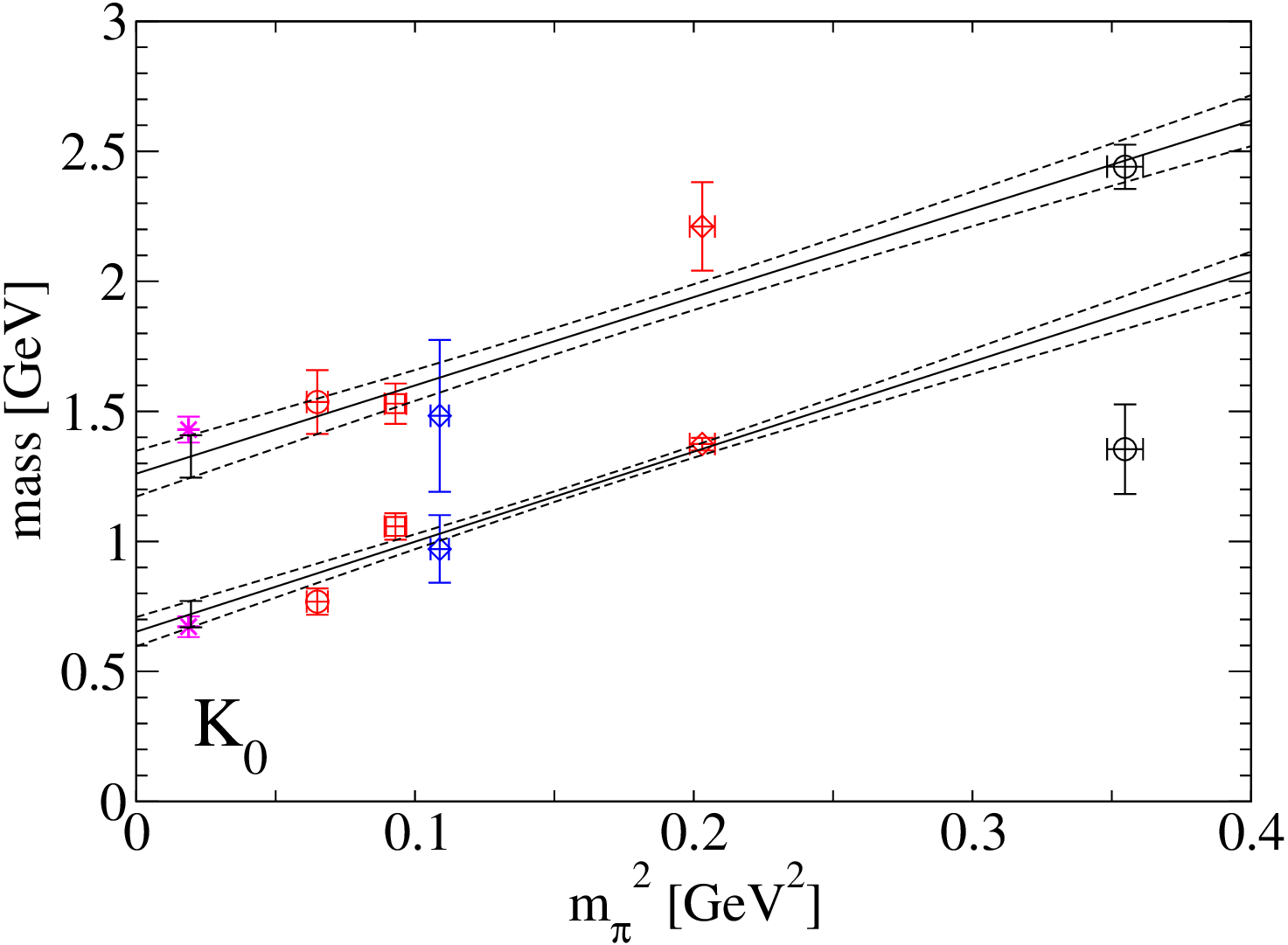}
\caption{
Mass plot for the strange 0$^{+}$ channel ($K_0^*$), ground state and first excitation.
}
\label{fig:K0*}
\end{minipage}
\end{figure}
\noindent
In the strange $0^{-{}}$ channel, interpolator (1) is used for the ground state and interpolators (1,2,8,17) for the excited state.
The extrapolated masses are close to the experimental ground state and the $K(1460)$  (see Fig.~\ref{fig:K}), 
confirming the latter, which is omitted from the summary table of  \cite{Nakamura:2010zzi}.

%%%%%%%%%%%%%%%%%%%%%%%%%%%%%%%%%%%%%%%%%%%%%%%%%%%%%%%%%%%%%%%%%%%%%%%%%%%%%%%
\subsection{The strange 0$^{+}$ channel: $K_0^*$}
\noindent
The $K_0^*(800)$ is a very broad resonance, omitted from the summary table of  \cite{Nakamura:2010zzi}.
We use the interpolators (10,12,13) in A50 and C72 and (4,10,12,13) in A66, B70 and C77, i.e., at small pion masses.
In B60 and C64 the signals are too noisy to produce a reliable value.
The chiral extrapolations are compatible with the $K_0^*(800)$ and the $K_0^*(1430)$ (see Fig.~\ref{fig:K0*}).
% As in the light scalar channel, we cannot exclude contributions from scattering states.

%%%%%%%%%%%%%%%%%%%%%%%%%%%%%%%%%%%%%%%%%%%%%%%%%%%%%%%%%%%%%%%%%%%%%%%%%%%%%%%
\subsection{The light 1$^{--}$ channel: $\rho$}
\begin{figure}
\begin{minipage}[t!]{74mm}
\includegraphics[width=73mm,clip]{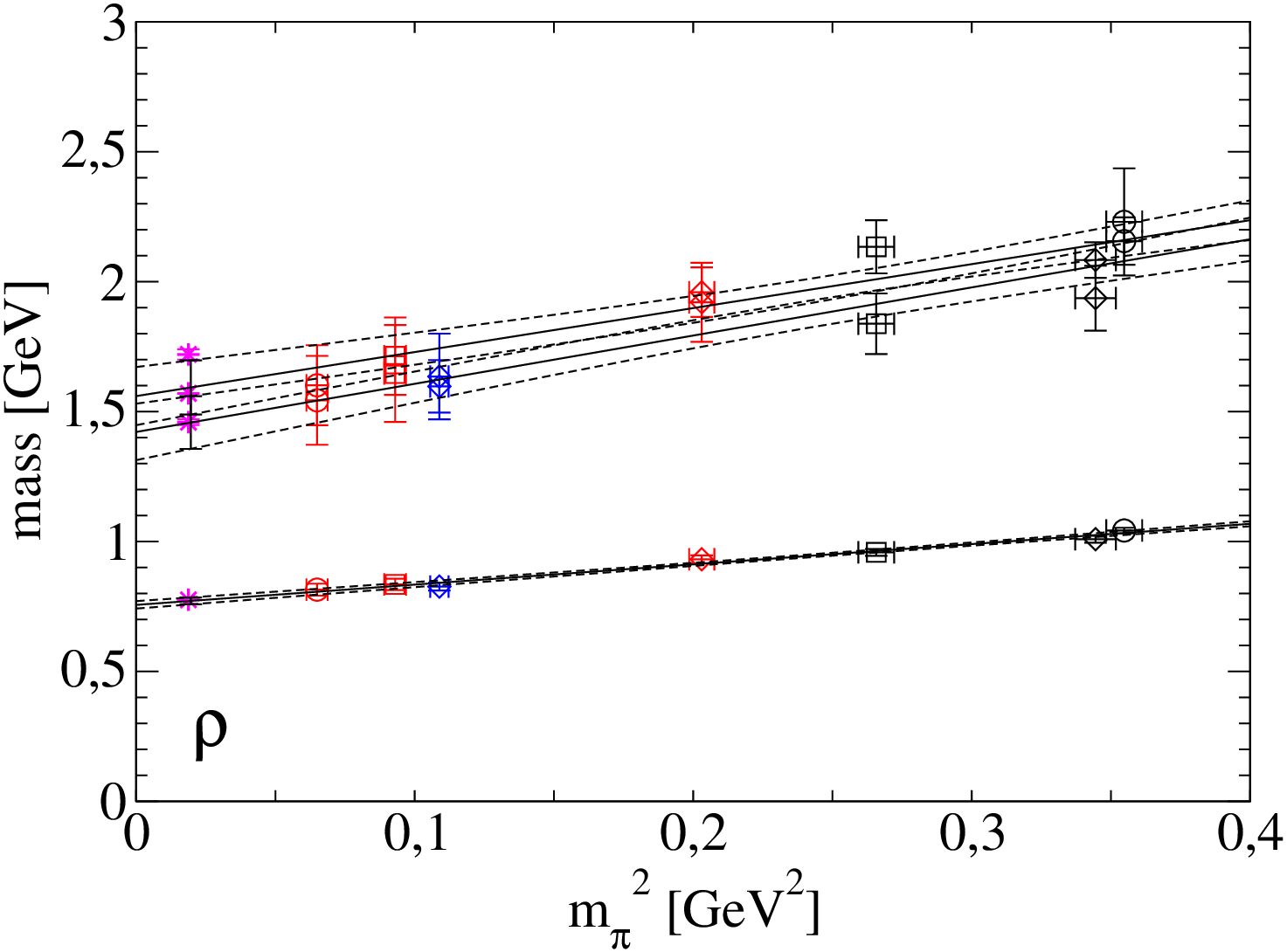}
\caption{
Mass plot for the 1$^{--}$ channel ($\rho$), ground state and two excitations.
}
\label{fig:rho}
\end{minipage}\hfill
\begin{minipage}[t!]{74mm}
\vspace*{-0.5mm}
\includegraphics[width=73mm,clip]{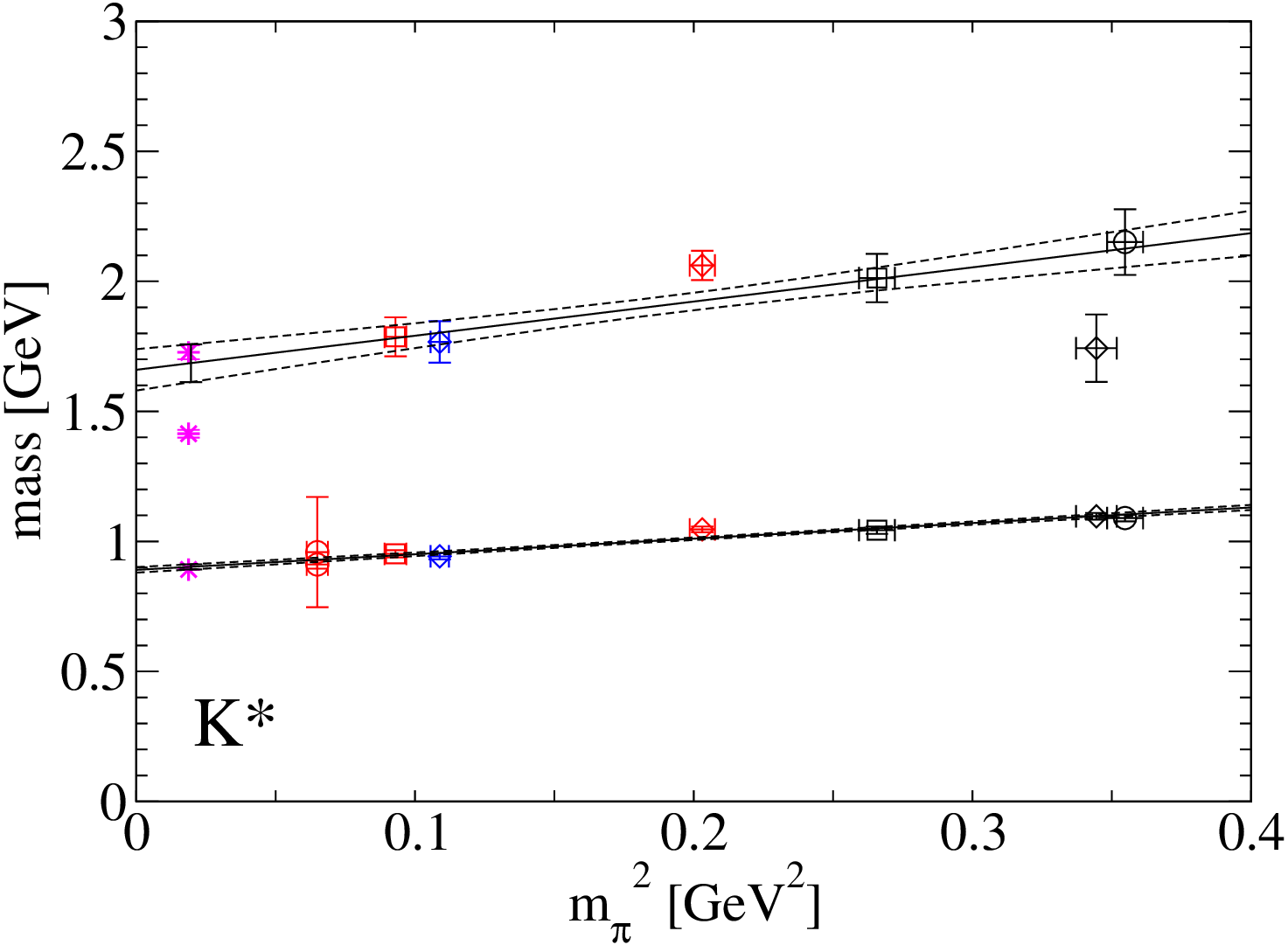}
\caption{
Mass plot for the strange 1$^{-}$ channel ($K^*$), ground state and first excitation.
}
\label{fig:K*}
\end{minipage}
\end{figure}
\noindent
The $\rho(770)$ comes out nicely as usual.
The excitations are extracted using subsets of the interpolators (1,8,12,17,18,20,22).
The excitations are very close (see Fig.~\ref{fig:rho}).
However, the eigenvectors do not confirm this picture.
Therefore, we extrapolate the results according to the na\"ively assumed level ordering.
The results are compatible with the experimental $\rho(1450)$ and $\rho$(1700 or 1570) within error bars.

%%%%%%%%%%%%%%%%%%%%%%%%%%%%%%%%%%%%%%%%%%%%%%%%%%%%%%%%%%%%%%%%%%%%%%%%%%%%%%%
\subsection{The strange 1$^{-}$ channel: $K^*$}
\noindent
In the strange $1^{-{}}$ channel, we use the set (1,8,9), which includes both types of approximate $C$-parities.
The ground state comes out nicely and is dominated by the $[C\approx-]$ interpolators.
For the first excitation, we find hardly any mixing at large pion masses, but some mixing in ensemble A66.
The corresponding A66 excited energy level is almost degenerate with the ground state, which remains a puzzle  (see Fig.~\ref{fig:K*}).
Due to its rather large error bar, the chiral extrapolation of the excitation still hits the $K^*(1680)$.
Our results suggest that simulations at smaller pion masses and with higher statistics are necessary in order to reliably describe the mixing of different $C$-parities and to be able to obtain the $K^*(1410)$.

%%%%%%%%%%%%%%%%%%%%%%%%%%%%%%%%%%%%%%%%%%%%%%%%%%%%%%%%%%%%%%%%%%%%%%%%%%%%%%%
\subsection{The light 1$^{++}$ channel: $a_1$}
\begin{figure}
\begin{minipage}[t!]{74mm}
\includegraphics[width=73mm,clip]{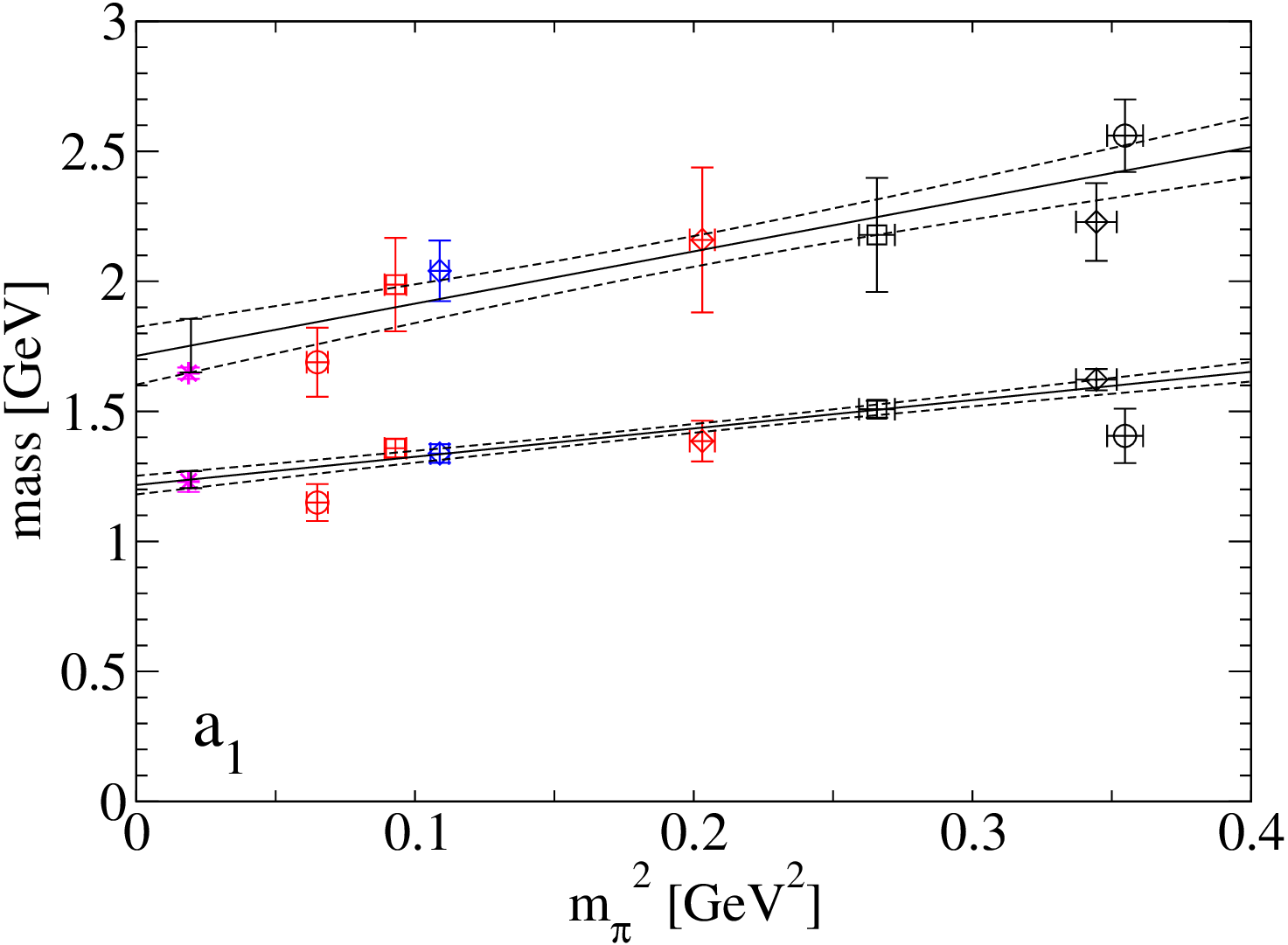}
\caption{
Mass plot for the 1$^{++}$ channel ($a_1$), ground state and first excitation.
}
\label{fig:a1}
\end{minipage}\hfill
\begin{minipage}[t!]{74mm}
\vspace*{-0.5mm}
\includegraphics[width=73mm,clip]{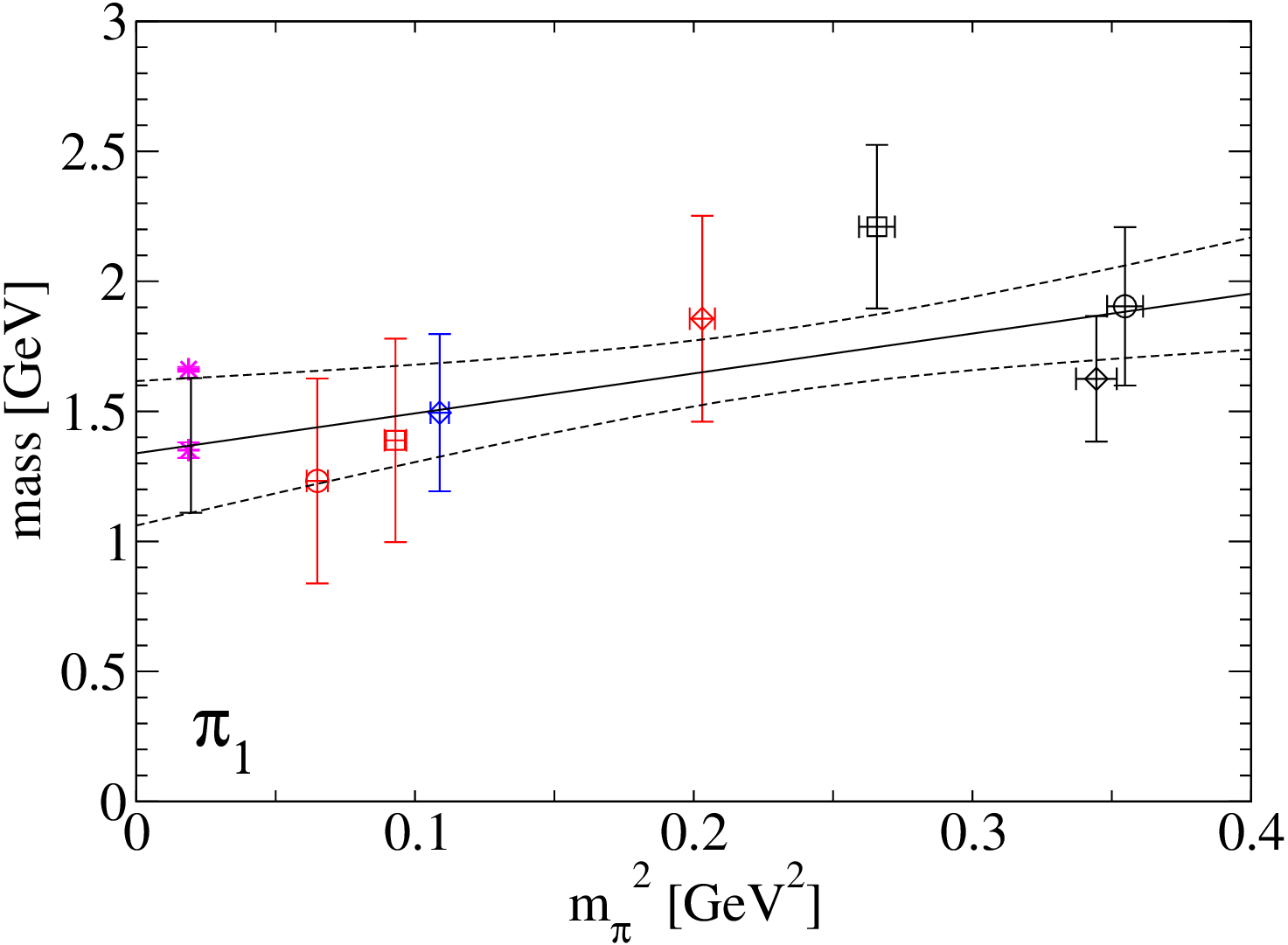}
\caption{
Mass plot for the exotic 1$^{-+}$ channel ($\pi_1$), ground state.
}
\label{fig:pi1}
\end{minipage}
\end{figure}
\noindent
In the 1$^{++}$ channel ($a_1$), the ground state is extracted using the single interpolator (1), 
the first excitation using subsets of (1,2,4,13,15).
Some of the plateaus drop slightly at large time separations. which may be a hint that the true ground state is not reached in our fit interval.
Nevertheless, the chiral extrapolations hit the experimental $a_1(1260)$ and the $a_1(1640)$ within error bars  (see Fig.~\ref{fig:a1}).
We thus confirm the existence of the latter, which is omitted from the summary table of  \cite{Nakamura:2010zzi}.

%%%%%%%%%%%%%%%%%%%%%%%%%%%%%%%%%%%%%%%%%%%%%%%%%%%%%%%%%%%%%%%%%%%%%%%%%%%%%%%
\subsection{The light 1$^{-+}$ channel: $\pi_1$}
\noindent
Due to the weak signal in the exotic $1^{-{+}}$ channel ($\pi_1$), the choice of operators is optimized in each ensemble separately.
Doing so, a mass can be extracted, albeit with comparatively large statistical uncertainty (see Fig.~\ref{fig:pi1}).
The chiral extrapolation hits the experimental $\pi_1(1400)$, but is also compatible with the $\pi_1(1600)$.

%%%%%%%%%%%%%%%%%%%%%%%%%%%%%%%%%%%%%%%%%%%%%%%%%%%%%%%%%%%%%%%%%%%%%%%%%%%%%%%
\subsection{The light 2$^{++}$ channel: $a_2$}
\begin{figure}
\begin{minipage}[t!]{74mm}
\includegraphics[width=73mm,clip]{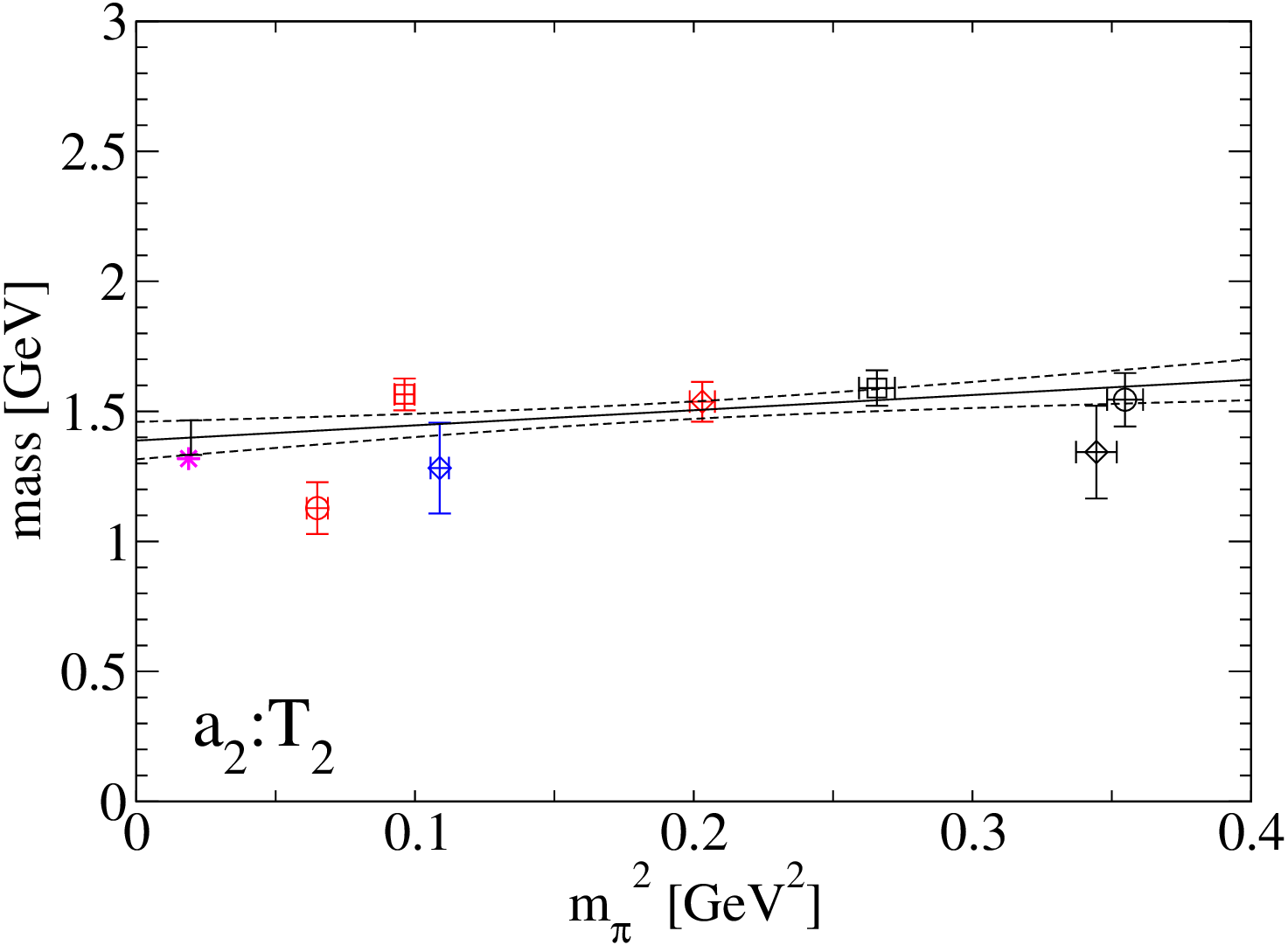}
\caption{
Mass plot for the 2$^{++}$ channel ($a_2$), ground state in representation T$_2$.
}
\label{fig:a2T2}
\end{minipage}\hfill
\begin{minipage}[t!]{74mm}
\vspace*{-0.5mm}
\includegraphics[width=73mm,clip]{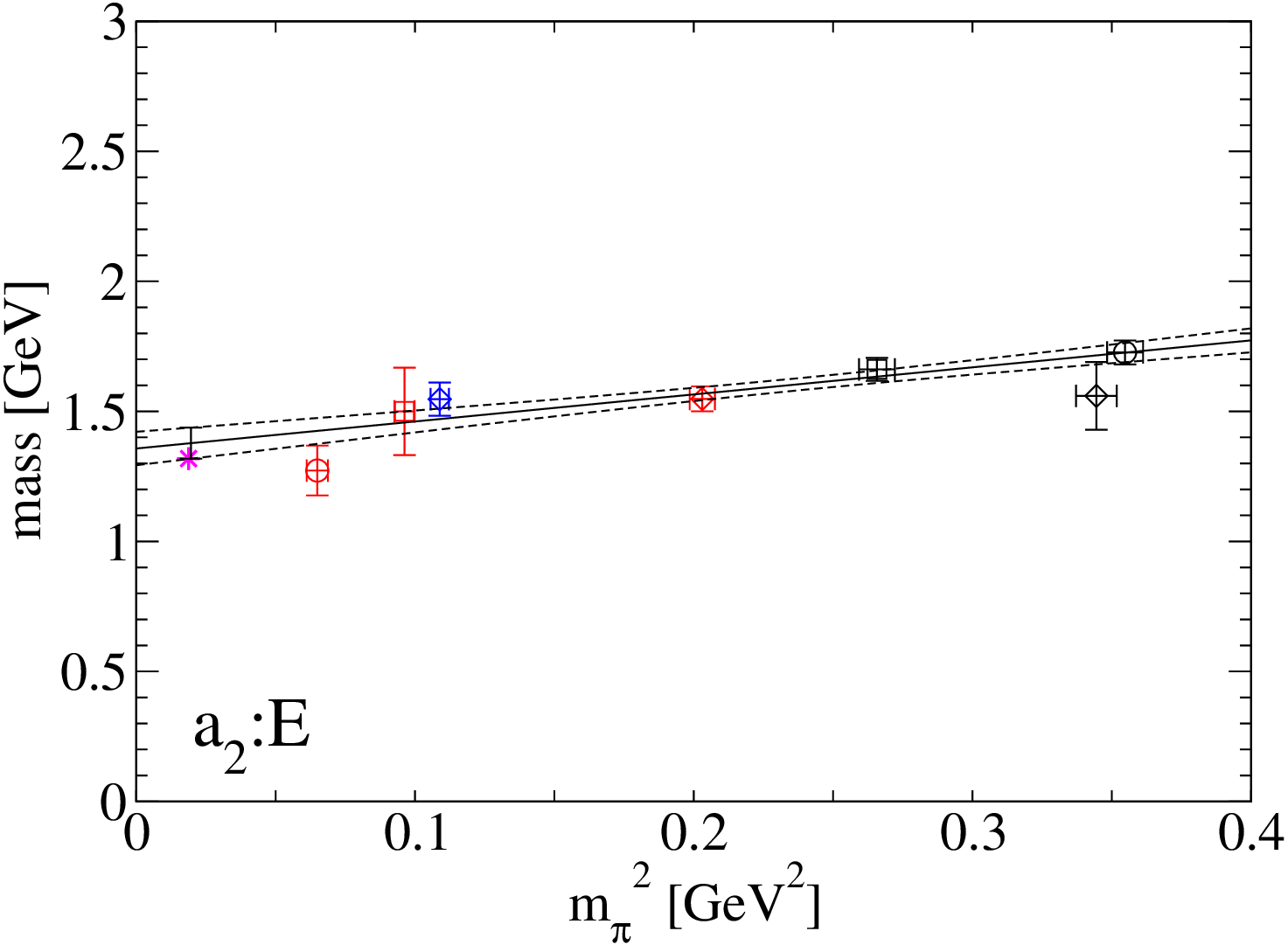}
\caption{
Mass plot for the 2$^{++}$ channel ($a_2$), ground state in representation E.
}
\label{fig:a2E}
\end{minipage}
\end{figure}
\noindent
In the $2^{+{+}}$ channel ($a_2$), interpolator (2) in $T_2$ and (2) (respectively (6) in A66) in $E$ is used to extract the ground state.
Some of the plateaus appear to be lighter than expected  (see Figs.~\ref{fig:a2T2} and \ref{fig:a2E}), 
however, the deviations being in a range which is compatible with statistical fluctuations.
The chiral extrapolations of the orthogonal lattice representations $T_2$ and $E$ agree 
and match the experimentally known resonance $a_2(1320)$ within error bars.

%%%%%%%%%%%%%%%%%%%%%%%%%%%%%%%%%%%%%%%%%%%%%%%%%%%%%%%%%%%%%%%%%%%%%%%%%%%%%%%
\section{Conclusions}
\label{sec:conclusion}
%%%%%%%%%%%%%%%%%%%%%%%%%%%%%%%%%%%%%%%%%%%%%%%%%%%%%%%%%%%%%%%%%%%%%%%%%%%%%%%
%
\begin{figure}[tbp]
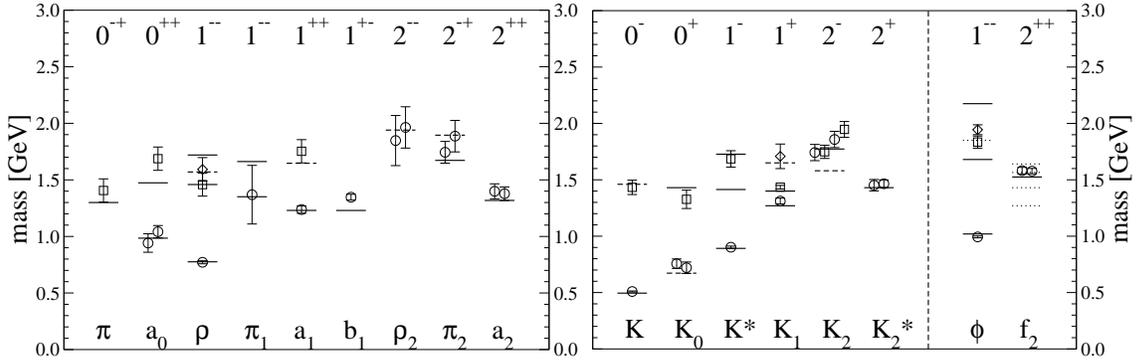

\begin{minipage}[h!]{74mm}
\includegraphics[width=76mm,clip]{collection_mesons.eps}
\end{minipage}
\hspace{2.2mm}
\begin{minipage}[h!]{74mm}
\includegraphics[width=72mm,clip]{collection_strange_mesons_yaxisright.eps}
\end{minipage}
\caption{
Collection of the mass results for light and strange mesons (left to right),  
obtained by chiral extrapolation of dynamical light quarks 
linear in the pion mass squared.
Experimental values listed by the Particle Data Group \cite{Nakamura:2010zzi} are denoted
by horizontal lines, dashed lines indicate needed confirmation.
Results shown side by side stem from different sets of
interpolators (resp.~$T_2$ and $E$ in spin 2). 
Strange quarks are analyzed in the partial quenching approximation.
Isoscalars suffer from neglected disconnected diagrams.
}
\label{fig:collection_masses}
\end{figure}
\noindent
We presented results for the excited meson spectrum from two dynamical Chirally Improved quarks.
Seven ensembles with pion masses in the range of 250 to 600 MeV have been analyzed.
Strange mesons have been treated in the partially quenched approximation.  
The majority of results agrees well with experiment (see Fig.~\ref{fig:collection_masses}), 
some states can be confirmed which are omitted from the summary table of the Particle Data Group \cite{Nakamura:2010zzi}.

%%%%%%%%%%%%%%%%%%%%%%%%%%%%%%%%%%%%%%%%%%%%%%%%%%%%%%%%%%%%%%%%%%%%%%%%%%%%%%%
\acknowledgments
%%%%%%%%%%%%%%%%%%%%%%%%%%%%%%%%%%%%%%%%%%%%%%%%%%%%%%%%%%%%%%%%%%%%%%%%%%%%%%%
\noindent
We thank C.~Gattringer, L.Y.~Glozman and S.~Prelovsek
for valuable discussions. The calculations have been performed on the SGI Altix
4700 of the Leibniz-Rechenzentrum Munich and on local clusters at ZID at the
University of Graz. We thank for this support. G.P.\,E., M.L. 
and D.M. have been supported by Austrian Science Fund (FWF): DK W1203-N16.
M.L.~has been supported by EU FP7 project HadronPhysics2. 
D.M.~acknowledges support by
COSY-FFE Projekt 41821486 (COSY-105) and by Natural Sciences and Engineering
Research Council of Canada (NSERC) and G.P.E., M.L.~and A.S.~acknowledge support by
the DFG project SFB/TR-55.

%\bibliographystyle{JHEP}
%\bibliography{/home/cbl/refs/Lgt}

\providecommand{\href}[2]{#2}\begingroup\raggedright\endgroup

\end{document}